# Electronic shells of Dirac Fermions in graphene quantum rings in a magnetic field

P. Potasz[1,2], A. D. Güçlü[1] and P. Hawrylak[1]

[1]*Institute for Microstructural Sciences, National Research Council of Canada, Ottawa, Canada*
[2]*Institute of Physics, Wroclaw University of Technology, Wroclaw, Poland*

**Abstract:**

We present results of tight binding calculations demonstrating existence of degenerate electronic shells of Dirac Fermions in narrow, charge neutral graphene quantum rings. We predict removal of degeneracy with finite magnetic field. We show, using a combination of tight binding and configuration interaction methods, that by filling a graphene ring with additional electrons this carbon based structure with half-filled shell acquires a finite magnetic moment.

## 1. Introduction

There is currently significant interest in graphene based electronics [1-6]. With a single sheet of graphene being a zero-gap semiconductor much effort is directed toward engineering gap in electronic spectrum of graphene by controlling its lateral size and shape [6]. In this approach edges of graphene nanoislands play an important role. The edge states are also associated with the possibility of magnetism in these carbon based nanostructures [7-9] and their effect can be maximized by constructing graphene quantum rings (GQR). Semiconductor quantum rings have already been developed to a level allowing the observation of Aharonov-Bohm effect of a single Fermion [10] but research on GQR is at an early stage. Noninteracting Dirac

Fermions in graphene quantum rings in a magnetic field have been investigated in the effective mass approximation [11, 12] and tight-binding models [13, 14], with preliminary results of AB oscillations in large rings reported experimentally [15]. Research on periodic arrays of GQR was also reported [16, 17].

In this work we present results of tight binding calculations demonstrating existence of degenerate electronic shells of Dirac Fermions in charge neutral graphene quantum rings and their evolution with perpendicular magnetic field. We then discuss the effect of filling of a degenerate shell with additional electrons. Using a combination of tight binding and configuration interaction methods we analyze ground state, spin and excitation spectrum of a half-filled shell. We show that the ground state does carry a finite total spin but the value of the spin does not correspond to either the maximum value predicted by the electron exchange (Hunds rule) nor the minimum value predicted by the noninteracting electron picture.

**2. The Model**

The building of a graphene ring investigated here is shown in Fig.1. It starts with a single benzene molecule around which other benzene molecules are attached, forming second and higher rings. This results in hexagonal graphene quantum dots. Next we remove layers from the interior, leaving a structure with a given size and width, with six inner and six outer well defined zigzag edges. Three of them are composed of A type and three of B type of non-equivalent carbon atoms of hexagonal non-Bravais carbon lattice. Inner and outer opposite edges are composed of two different types of atoms, similarly to zigzag nanoribbons [18-20].

**3. The single particle spectrum of a graphene quantum ring.**

We calculate the energy spectra of graphene quantum rings using the nearest-neighbor tight binding model. It has been successfully used for the calculations of electronic states of other low dimensional carbon-related materials such as nanotubes, nanoribbons or quantum dots with variety of the shape and size [6, 18-21]. The Hamiltonian is written as

$$H = \sum_i \varepsilon_i c_i^+ c_i + \sum_{<ij>} t_{ij} c_i^+ c_j + h.c. \tag{1}$$

, where $c_i^+, c_i$ are fermionic creation and anihilation operators on $i-th$ site, $\varepsilon_i$ are on-site energies, $t_{ij}$ are hopping matrix elements and $<ij>$ indicate summation over nearest neighboring sites (spin is not written explicitly). In this work, constant value for $t_{ij}$ is taken ($t = 2.5\,\text{eV}$), on-site energies $\varepsilon_i = 0$, and hydrogen passivation of the edge is assumed. If need be, modification of the hopping energies [22] or additional phases in the boundary conditions [23] can model passivation in the tight binding approach. The magnetic field perpendicular to graphene plane is incorporated in the hopping matrix elements using Peierls substitution as $t_{ij} \to t_{ij} \exp(i2\pi\phi_{ij})$, where $\phi_{ij}$ is Peierls phase given by $\phi_{ij} = \frac{e}{h}\int_{r_i}^{r_j} \mathbf{A}\mathrm{d}\mathbf{l} = B_z/2(x_i y_j - x_j y_i)$ with vector potential given in symmetric gauge $\mathbf{A} = B_z/2(-y, x, 0)$. In our calculations, we take $\phi_0 = h/e$ as the unit of the magnetic flux. We consider homogeneous magnetic field applied to the entire structure.

In Fig. 2 a single particle spectrum of one-benzene-ring thick structure consisting of N=96 carbon atoms is shown as a function of level index. Such a structure has 5 atoms on one of exterior edges and 3 atoms on one of the interior edges. 50% of atoms lie on the edges. Instead of a continuous Dirac spectrum we observe a gap at the Fermi level (Dirac point) and a grouping of states into quasi-degenerate shells. The lowest shell for electrons and for holes is six-fold degenerate. The probability density of three of these levels peaks mostly on the inner and the probability density of the other three levels mostly on the outer edge, as shown in the inset in Fig.2. The arrangement of outer edge levels forming shell is the following: in the first place lie two degenerate states (1, 2) and with a bit higher energy lies a single state (5). Between them lie two degenerate inner levels (3, 4). The state with highest energy corresponds to inner level (6). With increasing total number of atoms in a benzene thick ring, the degeneracy of shells is preserved and the number of shells increases (not shown here). The grouping of levels into two groups of three in each shell can be related to three outer edges of type A and three outer edges of type B atoms [24].

Fig. 3 shows the energy spectrum as a function of perpendicular magnetic field for a quantum ring from Fig. 2. The effect of the magnetic field is to remove degeneracy of a shell and introduce oscillations of shell energy levels periodic in magnetic field. The energy separation of different shells survives increasing magnetic field and changes smoothly with increasing magnetic field. This is in contrast to bulk graphene where Landau levels form.

**4. Magnetization of a half-filled shell.**

The main purpose of this paper is to determine the spin polarization of the electronic shell of graphene ring half-filled with additional electrons. One might expect these electrons to be spin polarized due to exchange interaction in analogy to graphene [7-9] and semiconductor quantum dots [25]. However, in this system, electron-electron interactions beyond exchange play major role due to high level of degeneracy. We calculate many body effects using a combination of tight binding and configuration interaction method. The many-electron Hamiltonian is given by

$$H = \sum_{s,\sigma} \varepsilon_s c^+_{s\sigma} c_{s\sigma} + \frac{1}{2} \sum_{\substack{s,p,d,f \\ \sigma,\sigma'}} \langle sp|V|df \rangle c^+_{s\sigma} c^+_{p\sigma'} c_{d\sigma'} c_{f\sigma} . \qquad (2)$$

Here, the first term corresponds to the single particle energies and molecular orbitals obtained by diagonalizing tight binding Hamiltonian, Eq.1, and shown in Fig.2. The second term describes interactions between the electrons added to the first electronic shell, as shown schematically in the upper-left hand corner of Fig.4. The two body Coulomb matrix elements $\langle sp|V|df \rangle$ are computed numerically using Slater-Koster $\Pi_z$ orbitals and screened by the effective dielectric constant κ, with values between 2 and 8 used in numerical calculations. In addition to the on-site interaction term, all scattering and exchange terms within next nearest neighbors, and all direct interaction terms are included. In our example of N=96 atom graphene ring with six-fold degenerate electronic shell, the half-filling corresponds to adding six electrons. With 6 electrons on 6-fold degenerate shell there are 400 configurations for total $S_z$=0. We build all configurations, construct a Hamiltonian matrix, diagonalize it and obtain energies and eigenfunctions of the interacting $N_{add}$=6 electron problem. By repeating computations with all allowed values of $S_z$ the total spin S is extracted from degeneracies of levels.

Fig. 4 shows the calculated energy spectra for different total spin S for half filled shell and effective dielectric constant κ=6. We find that the ground state has total spin S=1, separated by a very small gap from the total S=0 state. In exactly 6-fold degenerate system we expect total ground state spin S=3 due to the electron exchange (Hunds rule). In the noninteracting system we expect a nonmagnetic state S=0. Instead, in a correlated system configurations with two spin polarized electrons, shown schematically in the upper inset of Fig. 4, contribute to the ground state with finite total spin.

**5. Conclusions.**

We present results of tight binding calculations demonstrating existence of degenerate electronic shells in narrow, charge neutral graphene quantum rings, and using exact diagonalisation techniques predict finite total spin when graphene rings are filled with additional electrons.

Fig. 1. Construction of rings with different sizes from hexagonal structure. Red and blue atoms indicate presence of two equivalent sublattices in a honeycomb lattice. Circles indicate boundaries between consecutive layers. Atoms belonging to second layer are marked by different colors to improve transparency.

Fig 2. Energy spectrum near Fermi level and electronic charge densities of six levels forming one shell for ring structure consist of 96 atoms (5 atoms on the one outer edge). Three states (1, 2, 5) are localized mostly on outer and three (3, 4, 6) mostly on inner edges.

Fig. 3. Energy spectrum of graphene quantum ring shown in Figs.1,2 in a magnetic field.

Fig. 4. Ground and excited states of N=6 additional electrons on a six-fold degenerate shell (inset) as a function of total spin S. Ground state with S=1 and closely spaced excited state with S=0 are indicated.

Fig. 1

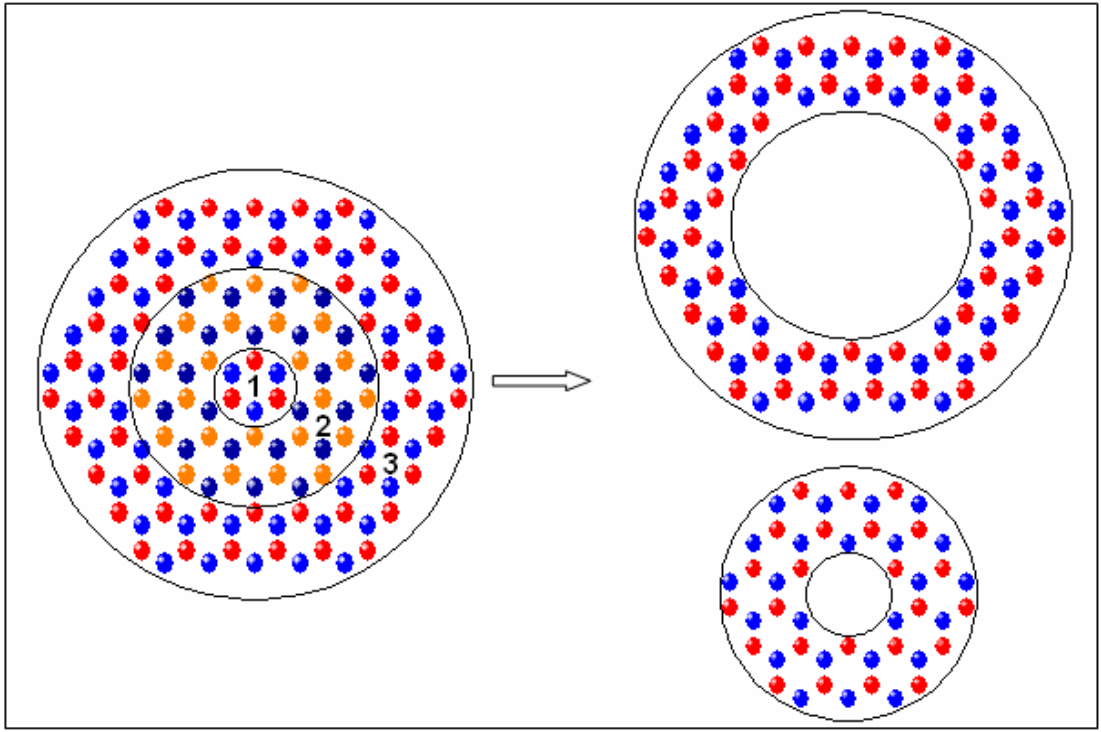

Fig. 2

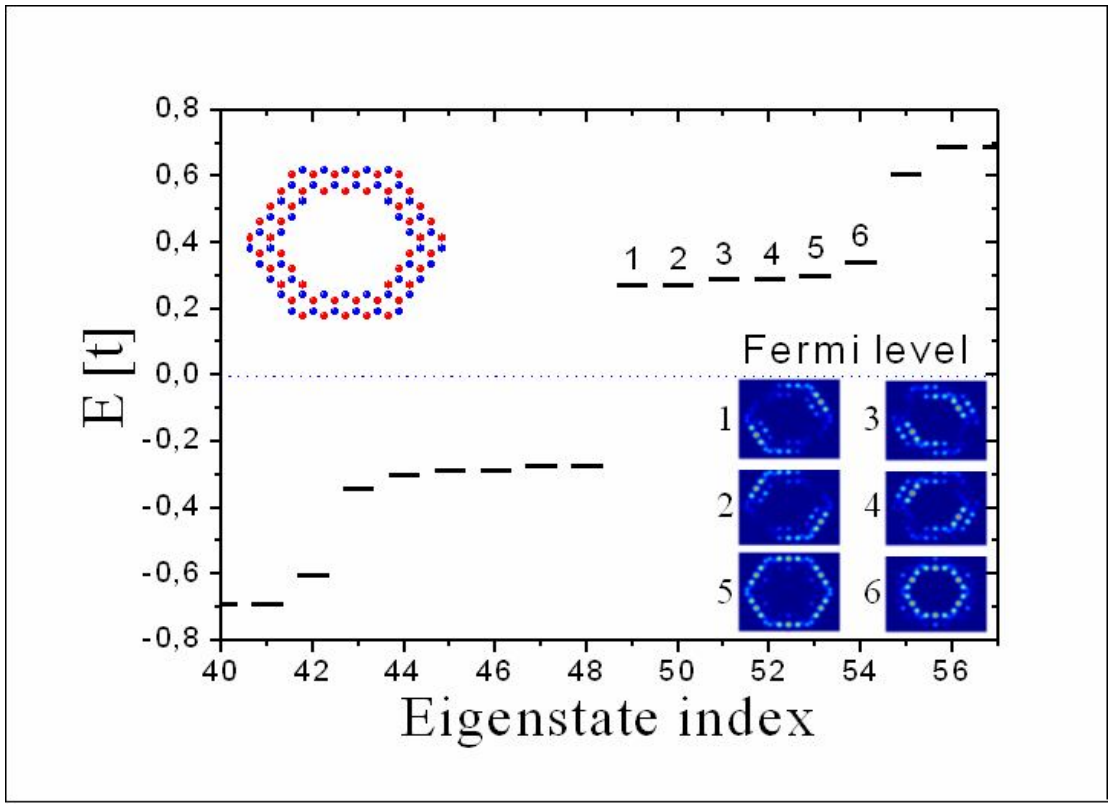

Fig. 3

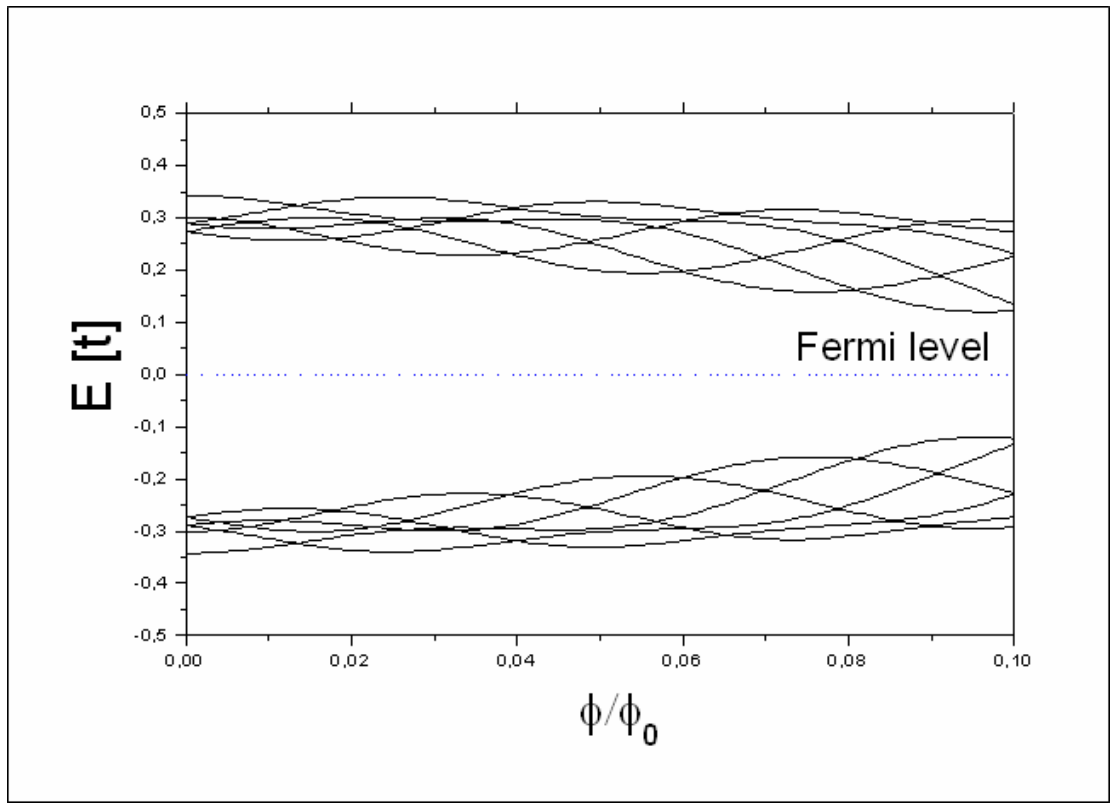

Fig. 4

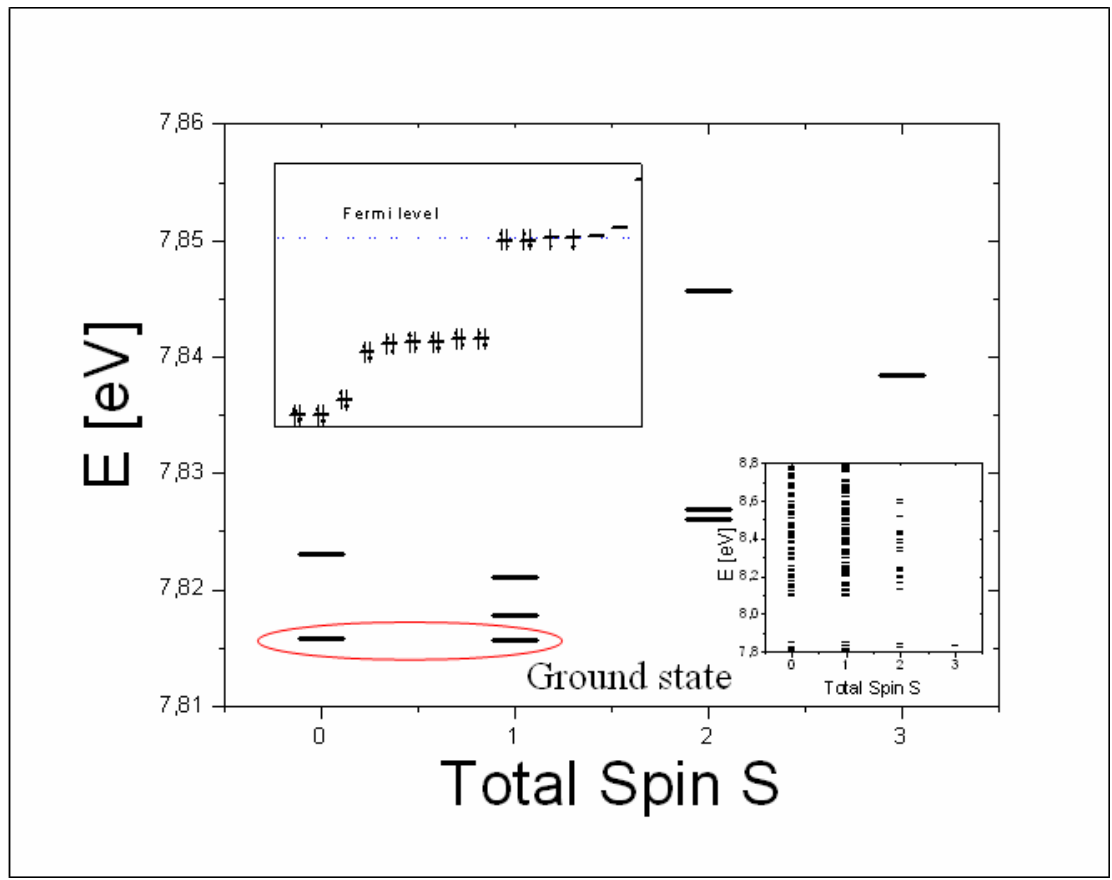